# ISO9000 BASED ADVANCED QUALITY APPROACH FOR CONTINUOUS IMPROVEMENT OF MANUFACTURING PROCESSES

**DEEB Salah – IUNG Benoît**

*Université Henri Poincaré UHP*
*Centre de Recherche en Automatique de Nancy (C.R.A.N) / UMR CNRS 7039*
*Faculté des Sciences - BP 239 - 54506 VANDOEUVRE (France)*
*Phone: +33 (0) 383 68 44 38 - Fax: +33 (0) 3 83 68 44 59*
*e-mail : salah.deeb@cran.uhp-nancy.fr*

Abstract: The continuous improvement in TQM is considered as the core value by which organisation could maintain a competitive edge. Several techniques and tools are known to support this core value but most of the time these techniques are informal and without modelling the interdependence between the core value and tools. Thus, technique formalisation is one of TQM challenges for increasing efficiency of quality process implementation. In that way, the paper proposes and experiments an advanced quality modelling approach based on meta-modelling the "process approach" as advocated by the standard ISO9000:2000. This meta-model allows formalising the interdependence between technique, tools and core value.

## 1. INTRODUCTION

The importance of Total Quality Management TQM has been considerably increased over the last years, on both practical and theoretical levels. As proposed by (Hellsten and Klefsjö, 2000) and exploited by (Hansson, and Klefsjö, 2003; Juan, and Vicente, 2004), the TQM has to be considered as a management system composed of three components which are the core values, the techniques and the tools. The three components are mutually dependant. Techniques and tools support the values and together they form a whole. Several core values (i.e. customer focus, leadership, continuous improvement, focus on process) and many techniques and tools (i.e. statistical process control (SPC), benchmarking, QFD, failure modes and effects analysis (FMEA)) have been already inventoried through different work (Sila, and Ebrahimpour, 2002; Dale, and McQuater, 1998). (Bunney, and Dale, 1997) highlighted that these tools and techniques are required for success of TQM, and the management must also consider these tools and techniques in order to advance towards total quality. Advance can be developed on different levels of the organisations such as the business and shop-floor levels (Oakland, 1989). For example, at the shop floor level, organisation is focusing on the continuous improvement of the manufacturing processes according to the business requirements as recommended in (IEC/ISO62264, 2002). Indeed in the last few years, many industrial organisations have looked upon continuous improvement in TQM as the mean by which they could maintain a competitive edge (Chin, et al., 2002) through a better control, at least, both of the product quality and process quality. This core value aims at reducing performance variability and increasing consistency in products, services and processes (Tummala and Tang, 1996). Many techniques and tools are known to support the core value of continuous improvement. In that way (Deming, 1986) proposed the PDCA-cycle as foundation of this concept. On this basis, (Geraedts, 2001) described that the ISO9000 quality system is quite suitable for seeking continuous improvement. (Johnson 2003) proposed also the PFMEA (Process Failure Mode and Effects Analysis) as a quality tool which support the practice and the philosophy of this value. This activity view of continuous improvement is extended by (Irani, et al., 2004) with innovation principle, which is seen as the successful exploitation of new ideas for this improvement.

Nevertheless, the main issue for performing the core value previously described, is the lack of formalisation of the technique used and the formalisation of the interdependence between the techniques and the two others components. Indeed techniques are most of the time informal with textual or graphical basis (Juan, and Vicente, 2004), without modelling of their interdependence (informal link). It ends in no optimisation of the efficiency of their implementations.

To face this issue, our proposal described in the paper, consists in developing an advanced quality modelling approach based on a meta-modelling (i.e., models of models, or models to create models (Mannarino, et al., 1997)) of the “process approach” advocated by the standard ISO9000:2000. This meta-model is completed by the integration of tools in terms of conventional quality/maintenance tool models and quality indicator models. This modelling approach is described in section 2. Then, section 3 is showing how this approach can be used in design and operation phases for continuous improvement of manufacturing process. The feasibility of this modelling approach is developed in section 4 for a lathe process. Finally, section 5 concludes the paper and gives direction for further work.

## 2. ADVANCED QUALITY APPROACH FOR CONTINUOUS IMPROVEMENT OF MANUFACTURING PROCESS

The advanced quality modelling approach formalises, firstly, a technique: a guide to be followed, secondly, the interdependence between this technique and some support tools (modelling of common objects) and finally the interdependence between this technique and the core value through the quality indicators model (assessment phase). The formalisation is based on the UML language (Unified Modelling Language) (Rumbaugh, et al., 1999) leading to model each concept of the technique and tools through class of objects and relationships between objects. Then this approach is computerised by means of an enterprise process modelling tool called MEGA Suite[1].

### *2.1 Meta-modelling of the technique*

The technique (or guide) proposed has to materialise a set of activities and rules to be performed in a logical order to carry out a continuous improvement of a manufacturing process. Indeed in design phase of the manufacturing process, this guide must help the quality engineer to better translate his “informal” requirements into formalised knowledge to better master the deployment of the manufacturing process.

Then the knowledge encapsulated in the resulting design model will be translated, in operation phase, to available information for the operator on site. This information allows to better control on line the quality drifts of the process or the product quality (implementation of the continuous improvement).

This technique is based on the meta-modelling of the “process approach” as defined in the standard ISO9000:2000 (AFNOR, 2000) and of some additional semantic concepts related to quality improvement (Dellea, *et al.,* 2002). Thus the translation of engineer requirements is done from informal expression to standardised quality entities such as product features, process feature, nonconformity …

The meta-modelling of the ISO9000:2000 standard is made on the following steps:

- Extraction, from the definitions enclosed in the standard, of all the entities (product, process, …) for identifying the quality concepts and their relationships (subject – verb- direct/indirect object),
- Modelling of each concept (entity) by a UML meta-class,
- Modelling of the links between the concepts through UML relationships which have a name and multiplicities,
- Modelling of the constraints between relationships or between meta-classes. These constraints materialise knowledge rules extracted from the textual definition of the concepts.

For example, from the ISO9000:2000 definition “Product is an output that results from a process”. The meta-modelling phase leads to identify two meta-classes and, one relationship with name and multiplicities (see figure 1).

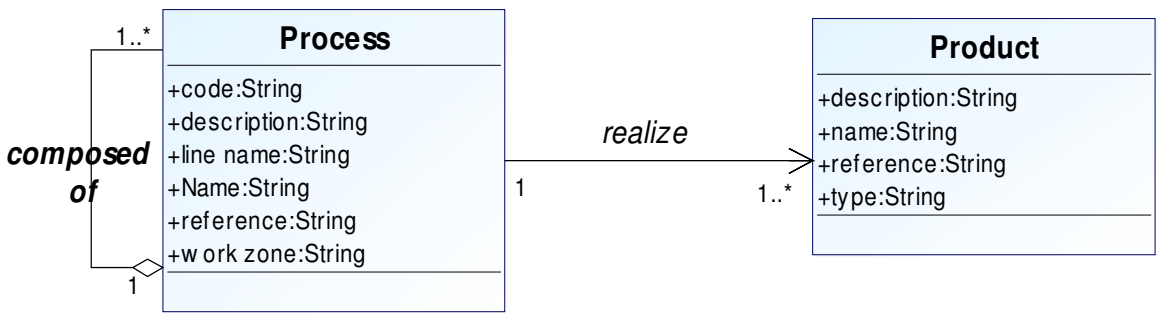

Fig. 1. Meta-modelling of the definition between the objects “product” and “process”.

The resulting meta-model is a diagram of classes of objects and relationships between theses objects (see figure 2).

The meta-model obtained from the ISO9000:2000 is improved by integrating quality modelling constructs based on “approved” principles such as the following ones:

- The general system theory already used in Manufacturing Engineering by (Mayer, et al., 1995). In this work, a product is mandatory defined by three attributes of shape, space and time, knowing that one manufacturing process transforms at least two of these attributes. Thus this principle lead to add the sub-types meta-classes “Time requirement”, “Space requirement” and “Shape requirement” to the meta-classes “Product requirement”.

[1] MEGA Suite, MEGA International, www.mega.com

- The maintenance standards as defined by IEC-50(191)[2]. The standards are meta-modelled by specialising sub-types related to the meta-class “Preventive action” in terms of “Schedule preventive action”, “Conditional preventive action” and “Predictive action”.
- The 5M principle proposed by (Ishikawa, 1963). It is meta-modelled by specialising the meta-class “Cause” with five sub-types: “Machine cause”, “Method cause”, “Material cause”, “Manpower cause” and “environMent cause” (see Figure 3).
The modelling constructs extend the semantic of the standard ISO9000:2000 meta-model.

*2.2 Interdependence between the technique and the quality tools.*

The second part of our advanced approach consists in meta-modelling the quality tools with regards to the technique. Indeed this meta-modelling has to support a way for formalising the interdependence between the technique and the tools in terms of common concepts (common entities). It allows from the technique, to “execute” a tool to better support the identification of the occurrence of the meta-classes and relationships. For example during the design phase, the use of FMEA tool facilitates the identification of the occurrences such as the causes of the nonconformity of the process, the action to be developed….

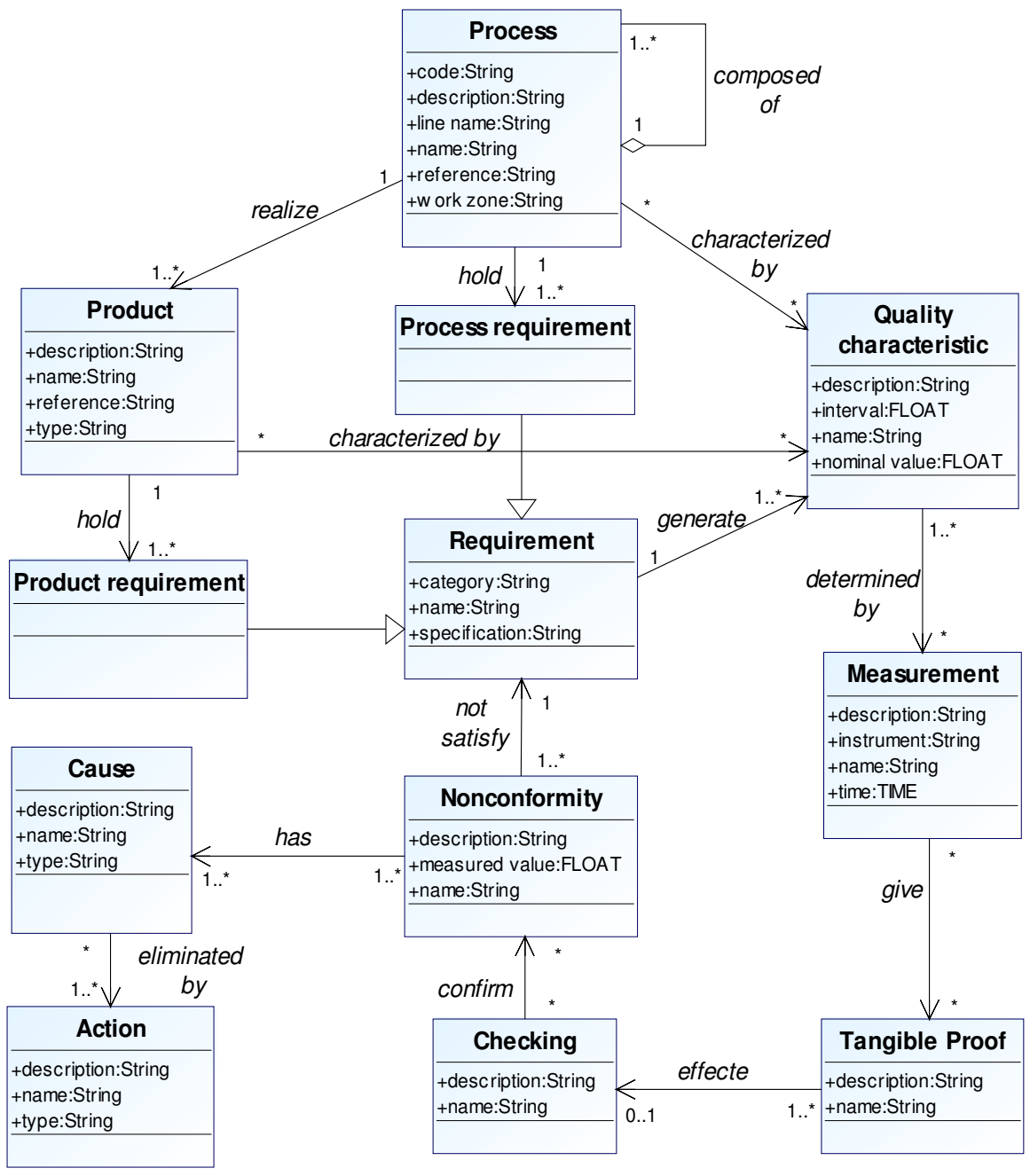

Fig. 2. Part of the meta-model of the standard ISO9000:2000.

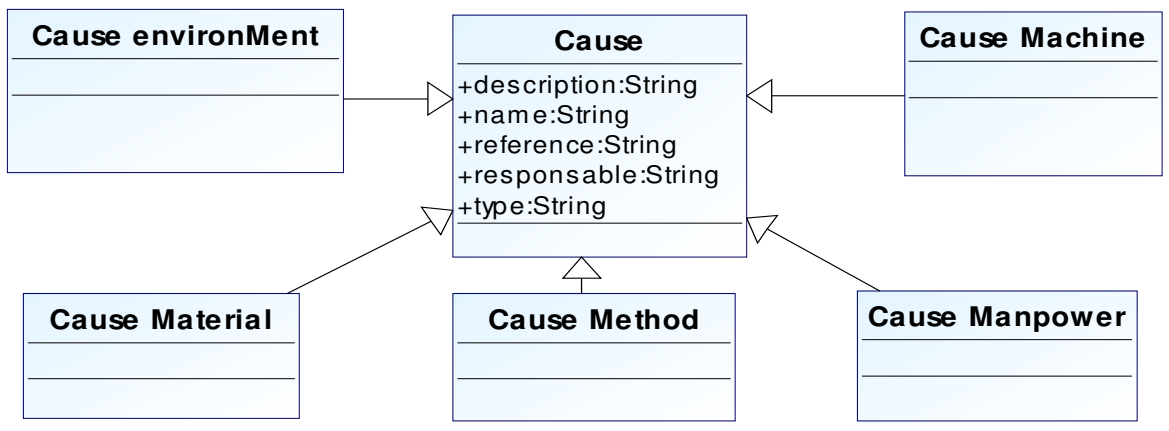

Fig. 3. The (5M) principle integrated to the meta-model of the standard ISO9000:2000.

At this moment, the technique (guide) integrates concretely two quality tools: Statistical Process Control (SPC) (AFNOR, 1996) and FMEA method IEC-60812[3].
It requires, for its deployment within the meta-model, to achieve the following actions:
- Modelling of each tool on the basis of its "textual" or normative description by developing the same steps as those described in (section 2.1) for ISO9000:2000 modelling.
- Integration of each tool model within the meta-model through the common objects between the different models (see figure 4).

*2.3 Interdependence between the technique and the core value*

The link between the technique and the selected core value (continuous improvement of manufacturing process) at shop floor level has to allow assessing the degree of fulfilment of this core value through adequate quality criteria.
These criteria are evaluated at the end of the design phase and can be, for example:
- A qualitative indicator to check the conformity and the nonconformity of the process or product. This indicator lies between 0 and 100%. It is calculated from the number of quality characteristics of process or product which are linked to a test, a measurement or an observation (quality concepts entities) knowing that each one of these controls is attached to a tangible proof. When all the quality characteristics of the process or product are checked, this indicator will take the value 100%; if not, the value will be included between 0 and 100%. In this last case, the core value will not be fully satisfied and the engineer could improve more his manufacturing process by adding other controls to the product/process requirements.
- A qualitative indicator for checking the cause existence in relation to the nonconformity of process or product. This indicator lies also between 0 and 100% and is calculated from the links connecting the meta-classes “Causes” and “Nonconformity”. If all nonconformities are connected at least to a cause, this indicator will take the value 100%. In the other cases, this indicator will oscillate between 0 and 100%. Thus the engineer, in the last case, could complete the lack of his knowledge related to the manufacturing process by identifying other causes of the nonconformity to increase qualitative level of his core value.

[2] IEC-50(191) International Electrotechnical Commission. vocabulary – chapter 191 « dependability and quality of service »

[3] IEC-60812 (41 pages) International Electrotechnical Commission – « Analysis techniques for system reliability - Procedure for failure mode and effects analysis (FMEA) ».

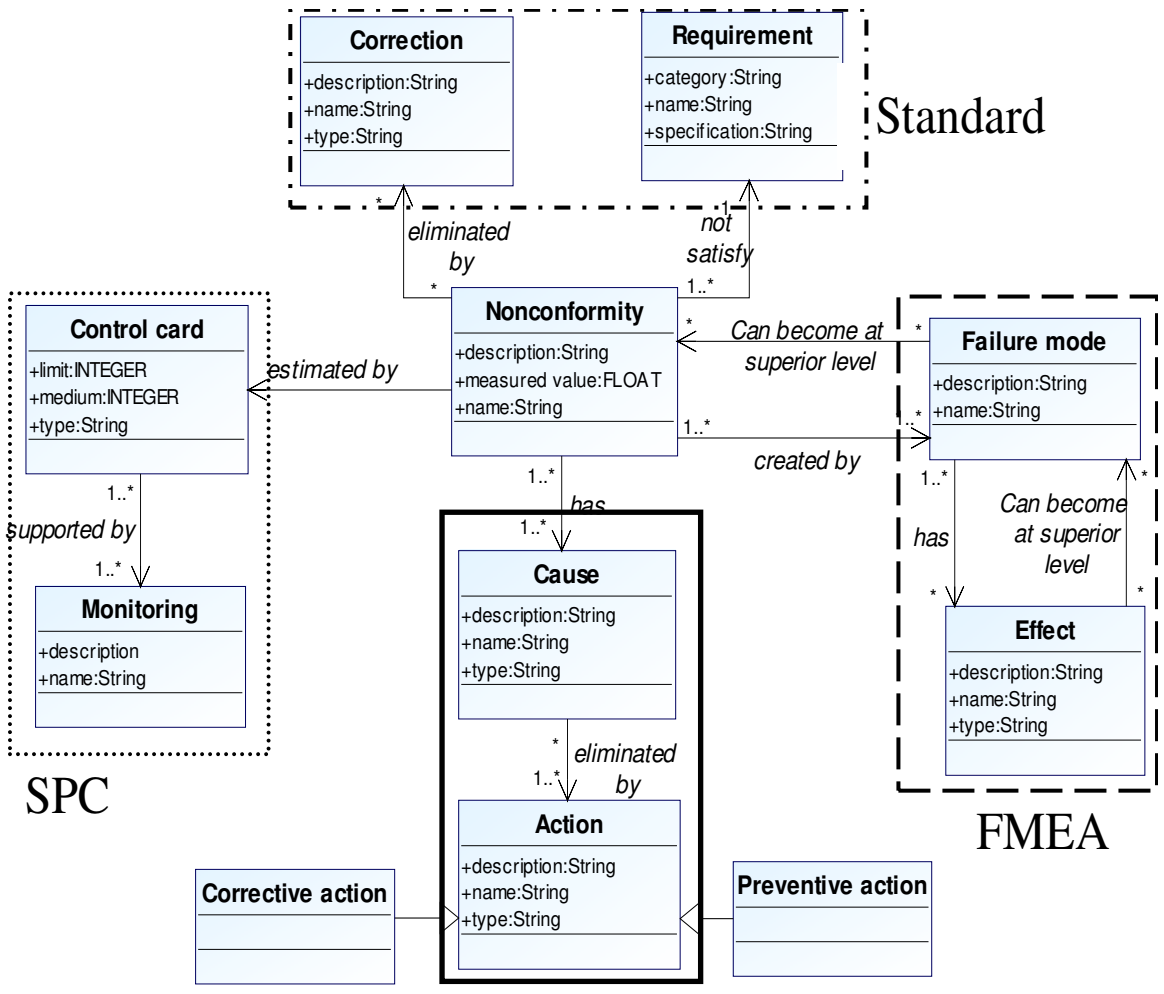

Fig. 4. Part of the integration of the two quality tools to the standard meta-model (common objects).

### *2.4 MEGA Suite for supporting the advanced quality approach*

To facilitate its use by quality specialists, this approach is supported by a CASE tool called MEGA Suite. MEGA Suite is an enterprise process modelling tool. It is based on its own meta-model which can be specialised for specific needs. The integration of our approach within MEGA is based on the following steps:

**1) Specialisation of the meta-model of MEGA**: The meta-model of MEGA is specialised with our needs. It means to embed our meta-model into a MEGA environment for creating a specific "MEGA environment". This "environment" will be used for developing specific models by instantiation procedures. It is composed of:

a) The meta-model of standard ISO9000:2000 (the technique), represented by quality process diagram.
b) The meta-models of quality tools, represented by quality tool diagrams.
c) Specific rules implemented into the quality process diagram. These rules are created using (modelling rule function) available in MEGA. These rules formalise the processing to be applied on the classes, relationships, constraints,… They ensure the consistency in the instantiation phase by guiding the engineer (section 3.1) in terms of question to be followed.
d) Control rules, implemented into quality entities (meta-class or relationship) of the quality process diagram (section 2.3) and performing quality level of the core value. Indeed, these rules defined in MEGA allow during the instantiation phase (section 3.1) to calculate the quality indicators (assessment phase).

**2) Database creation:**

a) Class diagram creation: This task consists in the implementation of the proposed meta-model (i.e. meta-model of standard ISO9000:2000 and quality tools integration) within MEGA using UML class diagram.
b) Database generation: MEGA suite enables automated derivation of the UML class diagrams in order to build up relational schema using Entity-Relationship (ER) diagram. Then, this relational schema generates a script which will be used to create the database. In the resulting database, meta-classes of the UML class diagram become tables and links between classes are automatically preserved.

In fact, this database will store the data resulting from the design phase (section 3.1) in order to use it during the operation phase (section 3.2).

## 3. USING APPROACH THROUGH MEGA SUITE

For the continuous improvement of a manufacturing process, the deployment through the MEGA Suite of the advanced quality modelling approach is based on two phases: design phase and operation phase.

### *3.1 In the design phase*

In this phase, the technique proposed leads to create a specific model related to the process studied and resulting from an **instantiation** phase of the quality process diagram. During this instantiation the quality engineer will be guided by all the formalised knowledge encapsulated in the meta-model and more precisely the rules coded in the quality process diagram of MEGA. These rules propose, under the form of questions, a coherent way to be followed in order to finalise the knowledge required to support a continuous improvement of the manufacturing process. For example, the creation of the instance "Process" related to the process studied will propose to the engineer to respect the relation (process – product) defined in the ISO9000:2000 standard and leading to create an instance to the "input product" entity and an instance to the "output product" entity. Then, the rules will propose to define the product requirements (shape, time and space requirements). So, this instantiation is based on seven steps materialising the chronological questions (each step associated with several questions): (1) *Definition of the context of the selected manufacturing process*. It concerns instantiation of the meta-classes "Product", "Process", "Customer" and "Supplier" and allows the decomposition of the process into elementary processes; (2) *Qualification and characterization of the product* (meta-classes "Requirement" & "Quality characteristic"); (3) *Identification of the means to determine the quality characteristics (*meta-classes "Observation", "Measurement" and "Test"); (4) *Identification of the means to determine the conformity or nonconformity* (meta-classes "Tangible proof", "Control", "Validation" and "Checking");

(5) *Identification of the nonconformities and conformities (*meta-classes "Nonconformity" and "Conformity"); (6) *Treatment of the causes of nonconformities (*meta-classes "Cause", "Preventive action" and "Corrective action"). It can lead also to execute the adequate quality tool for defining the right instances; (7) *Treatment of nonconformities (*meta-classes "Correction" and "Reject").

If the engineer does not have answers to all the questions asked during these 7 steps (incompleteness of his current knowledge), he can decide (a) to improve his knowledge leading to modification on the specific model (i.e. new instances) or (b) to continue the procedure but knowing that this "continuous improvement" is not the best one. This decision can be taken by using the result of the specific model assessment through the quality indicators model (section 2.3). This assessment is done by the MEGA rule (control rules) from the instance of the model.

### *3.2 In the operation phase*

At the end of the design phase, the knowledge encapsulated in the specific model (instance, attributes …) is stored in MEGA Suite within the database (section 2.4). This database is accessible through requests (Microsoft Access) in order to extract this knowledge and to put it on a support more accessible for the manufacturing operator. At this moment, we decided to translate the knowledge in Web page supports for providing information directly on site and on line through a PC closed the process. Thus the operator will be able to better control the quality drifts of the processes or product (implementation of the continuous improvement) by using the adequate information, for example:

- The potential causes of one process nonconformity observed by the operator on site. These causes can be shown under the form of cause tree, …
- The corrective and preventive actions to be developed in relation to the nonconformity observed and the potential causes identified previously …

## 4. APPLICATION CASE IN THE DESIGN PHASE

To show the feasibility of this modelling approach, a procedure of instantiation of a manufacturing process is in progress. Our case study is a lathe process used in our CIM centre AIPL-PRIMECA[4]. This manufacturing process transforms an input part (rough cylinder) into an output machined part which is then transferred to the following milling process. This lathe process is composed of several elementary processes like machining a cylinder with diameter "X", machining a cone,…

The current quality problems related to this lathe process is due firstly, to the drift of the values observed on the output part, and secondly, to the difficulty to make a coherent diagnosis that associates nonconformity of the output part with machine-tool cause. Each of the processes (global or elementary) could be considered as the starting point for the application of the seven instantiation steps (section 3.1). By this application, we focused on the "Machining a cylinder" process. The instantiation phase of this process is developed using the specific environment of MEGA shown in the section 2.4. The seven steps are planned through the rules implemented in the quality process diagram in order to guide us during the instantiation phase (see figure 5). In the following, the 1st and 2nd steps are presented. The first step allows creating "Machining a cylinder" as an instance of the meta-class "Process". The rules related to this step, guided us to create "Output part" as an instance of the meta-class "Product", "Transport by robot" as an instance of the meta-class "Supplier" and "Transport by carriage" as an instance of the meta-class "Customer". In the second step, the rules allowed to determine all the time, space and shape requirements of the "Output product". The process "Machining a cylinder" modifies the time and shape requirements but not the space one. Thus, we obtained the requirements "Cylindrical form of the turned part" as an instance of the meta-class "Shape requirement" and the requirement "manufacturing operation duration" as an instance of the meta-class "Time requirement". This step allows also instantiating the meta-class "Quality characteristic" by creating occurrences such as "Cylindrical part diameter, NV= 44.5 mm, IT= [-0.1, +0.1]" and "Manufacturing duration of the cylinder NV = 10 seconds IT = [-1, +1]", (NV: Nominal value; IT: Interval of Tolerance). Then, the instantiation phase continues through the last steps presented in the section 3.1.

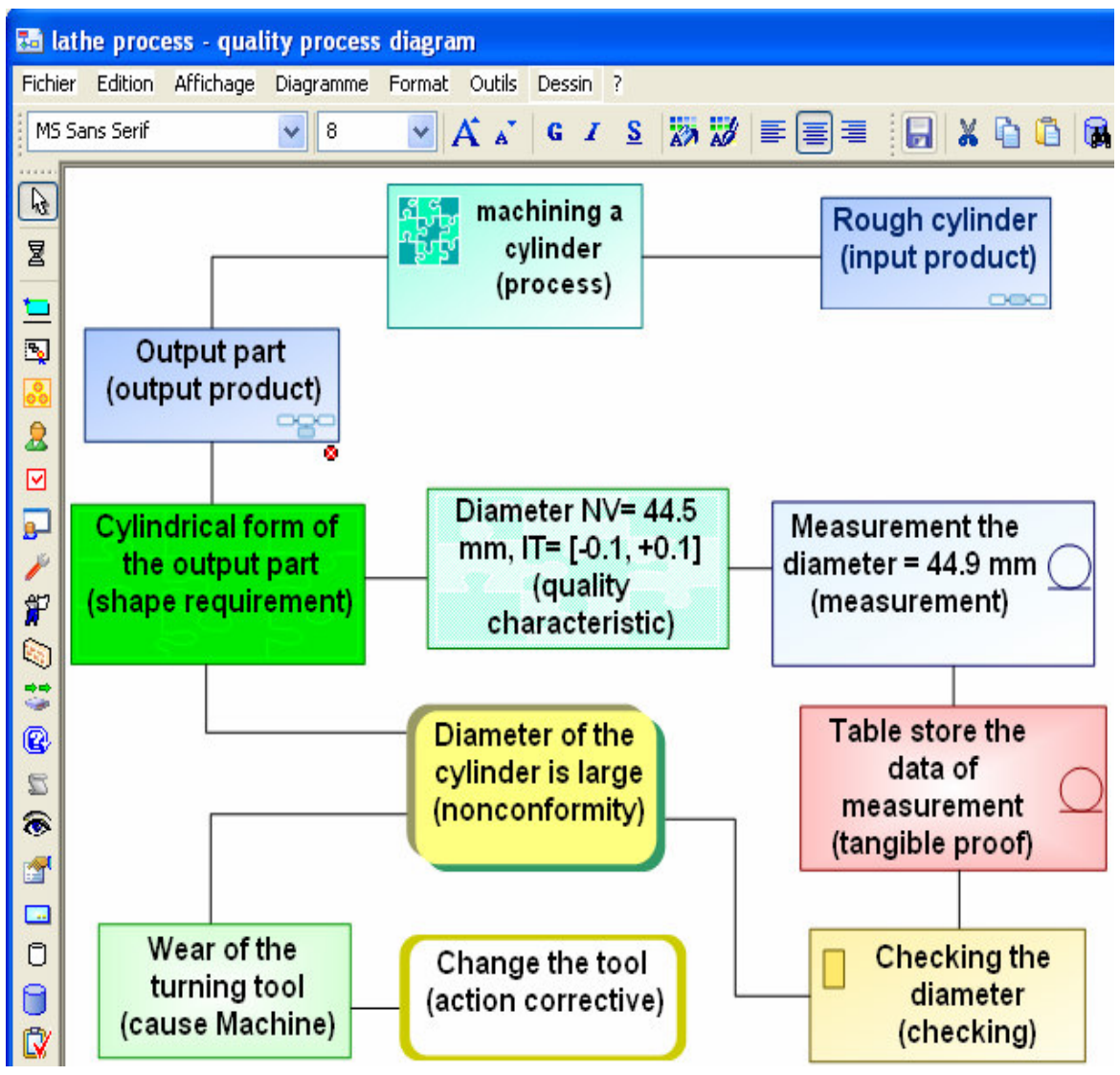

Fig. 5. Part of the specific model of the application case on MEGA using the specific environment.

Today, we can underline some results of this application for the design phase (operation phase is not yet supported):

- The "specialised MEGA tool" is able to support all the design phase related to a manufacturing process.

[4] http://www.aipl.uhp-nancy.fr/lorraine/

- The question supported by the rules allowed reconsidering deeply the observation means and the characterisation of the nonconformity.
- On the basis of the last point, the use of MEGA tool from the right entities, lead to a better identification of the relationships between causes and nonconformity.

## 5. CONCLUSION

In relation to TQM challenge, our main contribution consists in developing an advanced quality modelling approach for continuous improvement of manufacturing processes. This approach is based on a meta-modelling of the "process approach" advocated by the standard ISO9000:2000. This approach formalises, firstly, a technique: a guide to be followed, secondly, the interdependence between this technique and support tools (SPC and FMEA) and finally the interdependence between this technique and the core value (continuous improvement of manufacturing process) through the quality criteria (assessment phase). This approach is generic because it is based on standard and usable for many classes of applications resulting from different industrial fields. Many perspectives can be investigated:
- Extension of this approach toward a higher level of abstraction (management view).
- Integration into this approach of other quality indicators and other quality tools.
- Transformation of the "knowledge" encapsulated in the database to be available on site under the form of web page (operation phase).
- Testing of our approach on other applications cases and more precisely an industrial case.